\documentclass[reprint,
amsmath,amssymb,
aps,
]{revtex4-2}

\usepackage{graphicx}
\usepackage{dcolumn}

\begin{document}

\title{
Symmetry change of $d$-wave superconductivity\\
in $\kappa$-type organic superconductors
}

\author{
S. Imajo$^{1,2,*},\thanks{imajo@issp.u-tokyo.ac.jp}$
K. Kindo$^2$,
and
Y. Nakazawa$^1$
}
\affiliation{
$^1$Graduate School of Science, Osaka University, Toyonaka, Osaka 560-0043, Japan\\
$^2$The Institute for Solid State Physics, the University of Tokyo, Kashiwa, Chiba 277-8581, Japan
}

\date{\today}

\begin{abstract}
 Magnetic-field-angle-resolved heat capacity of dimer-Mott organic superconductors $\kappa$-(BEDT-TTF)$_2$$X$ is reported.
Temperature and field dependence of heat capacity indicates that the superconductivity has line nodes on the Fermi surface, which is a typical feature of $d$-wave superconductivity.
In-plane field-angle dependence exhibits fourfold oscillations as well as twofold oscillations due to anisotropy of superconducting gap functions.
From analyses of the fourfold term, the gap symmetry is determined as $d_{x^{2}-y^{2}}$+$s_{\pm}$ for $X$=Cu[N(CN)$_2$]Br and $d_{xy}$ for $X$=Ag(CN)$_2$H$_2$O.
We suggest that the symmetry difference comes from the competition of $d_{xy}$ and $d_{x^{2}-y^{2}}$ antiferromagnetic fluctuations depending on lattice geometry.
\end{abstract}

\maketitle
\begin{figure}
\begin{center}
\includegraphics[width=\hsize]{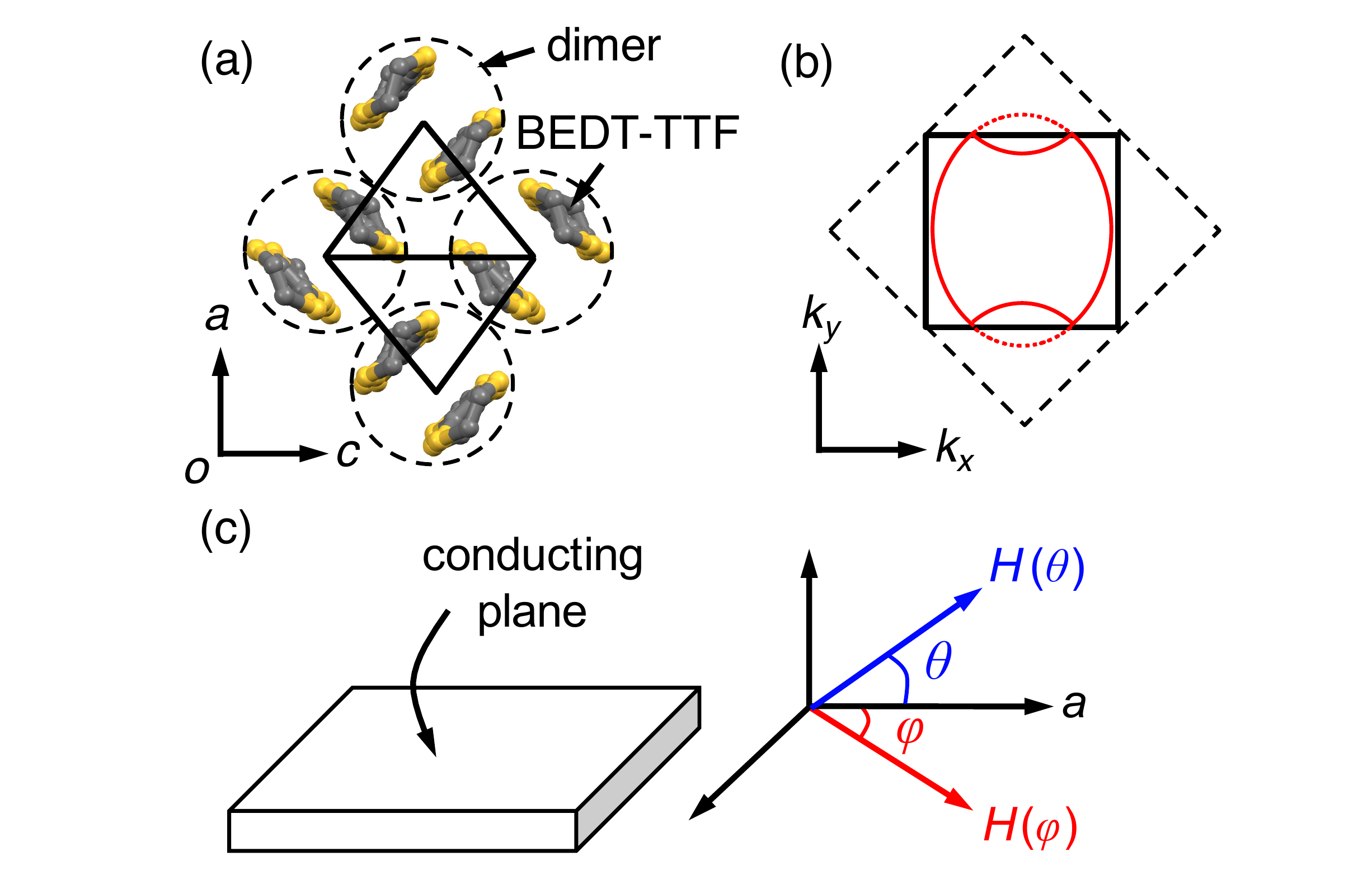}
\end{center}
\caption{
(a) Molecular arrangement of BEDT-TTF in the conducting plane of $\kappa$-(BEDT-TTF)$_2$Cu[N(CN)$_2$]Br.
(b) Fermi surface of $\kappa$-(BEDT-TTF)$_2$$X$ (red) in the first Brillouin zone (solid lines) and the extended Brillouin zone (dashed lines).
As indicated, the axes $k_x$ and $k_y$ are defined according to the axes of the first Brillouin zone ($k_b$, $k_a$) for $\kappa$-Ag and ($k_a$, $k_c$) for $\kappa$-Br, respectively.
(c)The definition of magnetic field directions against the two-dimensional conducting plane of the sample.
The polar angle $\theta$ signifies the inclination from the plane while the azimuthal angle $\phi$ denotes the in-plane angle from the $a$ axis.
}
\label{fig1}
\end{figure}

  $\kappa$-(BEDT-TTF)$_2$$X$ organic conductors (BEDT-TTF and $X$ represent bis(ethylenedithio)tetrathiafulvalene and a monovalent anion, respectively) have been extensively studied as proto-type dimer-Mott compounds.
These salts have a quasi-two-dimensional (quasi-2D) electronic system arising from the alternating layered structure of the organic donor BEDT-TTF and the counter anion $X$.
As shown in Fig.~\ref{fig1}(a), the BEDT-TTF molecules form dimer units that are arranged in the distorted triangular lattice structure.
Depending on the lattice geometry and the ratio of the inter/intra dimer transfers, the ground states of $\kappa$-type compounds vary between metal, superconductivity, and antiferromagnetic insulator as well as quantum spin liquid.
In a typical electronic phase diagram of $\kappa$-type salts\cite{1}, a superconducting phase neighbors an antiferromagnetic Mott insulating phase.
The appearance of the superconducting phase by suppressing the antiferromagnetic state with the application of external/chemical pressures implies that the origin of the superconductivity is closely related to the antiferromagnetic spin fluctuations enhanced at the verge of the antiferromagnetic phase.
When the superconductivity is mediated by the antiferromagnetic fluctuations, the superconducting energy gap function is anticipated to be anisotropic in the momentum space.
In the case of well-known high-$T_c$ cuprates whose superconductivity presumably arises from the antiferromagnetic spin fluctuations, their square lattice structure and almost 1/2 filling of the energy band make the $d_{x^{2}-y^{2}}$-wave superconductivity stable.
Based on a situation similar to cuprates, $\kappa$-(BEDT-TTF)$_2$$X$ would have $d$-wave symmetry, or, more specifically, $d_{xy}$-wave, symmetry in the extended Brillouin zone for the unfolded one-band dimer model, as shown by the outer dashed lines in Fig.~\ref{fig1}(b).
For organics, however, it is necessary to consider whether the strength of the dimerization is sufficient to apply the dimerization approximation or not.
Also, the effect that the dimers form the distorted triangular lattice must be taken into account.
Numerous experimental works for determining its pairing symmetry have been reported.
However, three contradictory conclusions have been suggested: $s$-wave\cite{2,3,4}, $d_{x^{2}-y^{2}}$-wave (=$d_{x^{2}-y^{2}}$+$s_{\pm}$-wave\cite{4.5})\cite{5,6,7,8}, and $d_{xy}$-wave \cite{9,10} symmetry.
Theoretical studies have also proposed $d_{x^{2}-y^{2}}$+$s_{\pm}$-\cite{11} and $d_{xy}$-wave\cite{12,13,14} symmetry depending on their models.
Some recent theories based on a more realistic model\cite{15,16,17,18} suggest that the two symmetries compete and the emergent symmetry is determined according to the ratio of transfer integrals.
Nevertheless, the experimental evidence of the competition of the emergent symmetry is still absent.

 To discuss the pairing symmetry of the dimer-Mott organic superconductors, in this work, we determine the superconducting gap structure of $\kappa$-(BEDT-TTF)$_2$Ag(CN)$_2$H$_2$O and $\kappa$-(BEDT-TTF)$_2$Cu[N(CN)$_2$]Br (hereafter, we abbreviate them as $\kappa$-Ag and $\kappa$-Br, respectively) by employing angle-resolved heat capacity measurements capable of detecting the anisotropy of the gap functions\cite{19,20,21}.
From the analyses of quasiparticle excitation over the superconducting gap, we classify the superconductivity into quasi-2D line-nodal superconductivity.
However, the detected anisotropy suggests that the detailed symmetry of the two salts is given as $d_{xy}$ for $\kappa$-Ag and$d_{x^{2}-y^{2}}$+$s_{\pm}$ for $\kappa$-Br.

 The single crystals measured in this study were synthesized by electrochemical oxidation of BEDT-TTF donors with electrolytes of counter anions.
Heat capacity measurements were performed by using a high-resolution relaxation calorimeter\cite{22} in rotating magnetic fields similar to the reported method\cite{23} and by using a calorimeter for pulsed magnetic fields\cite{24} in a $^{3}$He cryostat which can cool the samples down to $\sim$0.6~K.
We confirmed that there is no angle dependence of background heat capacity.
The samples were mounted on the calorimeter according to crystal axes determined by x-ray diffraction in advance.
The field angle was determined by the Hall voltage signal of a Hall device attached to the sample holder.

\begin{figure}
\begin{center}
\includegraphics[width=\hsize]{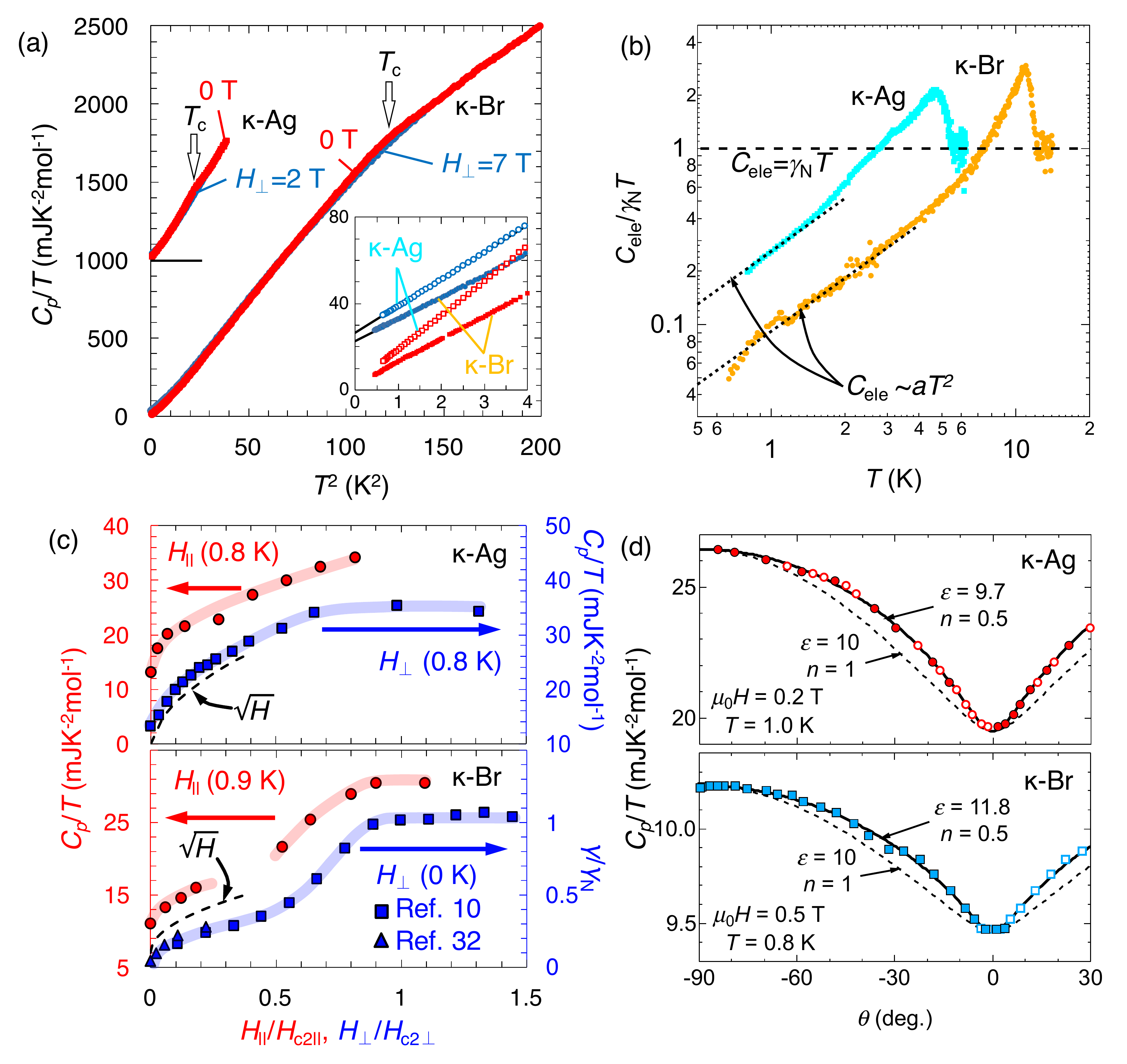}
\end{center}
\caption{
(a) $C_p$/$T$ of $\kappa$-Ag and $\kappa$-Br as a function of squared temperature $T^{\rm 2}$.
For clarity, an offset is added to the data of $\kappa$-Ag.
The red squares denote the 0 T data, while the blue circles represent the data of the normal state in magnetic fields perpendicular to the conducting plane $H_{\perp}$.
The inset is an enlarged graph of (a) below $T^2$=4~K$^2$.
The black solid lines are the linear fit to the data of the normal state.
The open and solid symbols denote the data for $\kappa$-Ag and $\kappa$-Br, respectively.
(b) Temperature dependence of the superconducting electronic heat capacity $C_{\rm ele}$/$\gamma_{\rm N}$$T$.
The dashed line indicates the electronic heat capacity of the normal state.
The dotted lines indicate the quadratic temperature dependence of $C_{\rm ele}$.
(c) Recovery of the low-temperature electronic heat capacity in fields.
The top and bottom panels show the data for $\kappa$-Ag and $\kappa$-Br, respectively.
The red circles (the left scale) represent the data in parallel fields whereas the blue squares (the right scale) indicate those in perpendicular fields.
The data of $\kappa$-Br are taken from Ref.~\cite{24} (red circles), Ref.~\cite{9} (blue squares), and Ref.~\cite{30} (blue triangles).
The black dashed curves signify the $\sqrt{H}$ dependence, and the colored thick curves are guides for the eye.
(d) Polar-angle $\theta$ dependence of heat capacity of $\kappa$-Ag and $\kappa$-Br.
The solid and open symbols represent the obtained data and their mirrored reflection at $\theta$=0~$^{\circ}$ to confirm symmetric features.
The solid and dashed curves are the fits to Eq.~(1) with the displayed parameters $\epsilon$ and $n$.
}
\label{fig2}
\end{figure}

 The temperature dependence of the heat capacity is presented in a $C_p$/$T$ vs $T^2$ plot in Fig.~\ref{fig2}(a).
As indicated by the arrow, the anomalies associated with the superconducting transition are observed at $\sim$5.2~K for $\kappa$-Ag and $\sim$11.3~K for $\kappa$-Br.
Despite the large phonon contribution coming from the soft lattice of the organics, the anomalies can be seen in this plot.
The inset enlarges the plot below $T^2$=4~K$^2$.
The electronic heat capacity coefficient $\gamma_{\rm N}$ is estimated from the fitting of the data of the normal state to $C_p$/$T$=$\gamma_{\rm N}$+$\beta$$T^{\rm 2}$ with the lattice heat capacity coefficient $\beta$.
The values of $\gamma_{\rm N}$ and $\beta$ are given as 26.6$\pm$0.3~mJK$^{-2}$mol$^{-1}$ and 12.4$\pm$0.2~mJK$^{-4}$mol$^{-1}$ for $\kappa$-Ag and 22.5$\pm$0.2~mJK$^{-2}$mol$^{-1}$ and 10.0$\pm$0.2~mJK$^{-4}$mol$^{-1}$ for $\kappa$-Br, respectively.
These values agree with the reported values\cite{9,25,26,27,28,29}.
The scaled electronic heat capacity $C_{\rm ele}$/$\gamma_{\rm N}$$T$ at 0~T is shown as a function of temperature in Fig.~\ref{fig2}(b).
It is obtained by subtracting the lattice heat capacity estimated from the data of the normal state above $H_{\rm c2}$, which is described by the total of the lattice heat capacity and $\gamma_{\rm N}$$T$.
As reported in earlier studies\cite{25,26,27,28,29,30}, the low-temperature $C_{\rm ele}$ shows $\sim$$aT^2$ behavior, which is evidence that line nodes exist in the two-dimensional Fermi surface.
The coefficient of the quadratic term $a$ is obtained as 7.4~mJK$^{-3}$mol$^{-1}$ for $\kappa$-Ag and 2.0 mJK$^{-3}$mol$^{-1}$ for $\kappa$-Br.
Assuming the gap function is a simple $d$ wave, it is known that the coefficient is expected to be $\sim$3.3$k_{\rm B}$$\gamma_{\rm N}$/$\Delta$$_0$\cite{28,30}, where $\Delta$$_0$ is the gap amplitude at 0~K.
This equation gives $\Delta$$_0$ for $\kappa$-Ag and $\kappa$-Br as 2.3$k_{\rm B}$$T_{\rm c}$ and 3.2$k_{\rm B}$$T_{\rm c}$, respectively.
Comparing with $\Delta$$_0$=2.14$k_{\rm B}$$T_{\rm c}$ for weak-coupling $d$-wave superconductors, it is found that $\kappa$-Ag is in a weak-coupling regime while the Cooper pairing of $\kappa$-Br is rather strong.
Figure~\ref{fig2}(c) presents $C_p$/$T$ as a function of external field reduced by $H_{\rm c2}$ in parallel and perpendicular directions\cite{29,29.5} with the reported $\gamma$/$\gamma_{\rm N}$ of $\kappa$-Br in perpendicular fields (the blue squares and triangles in the bottom panel)\cite{28,30}.
In the low-field region, the data clearly show $\sqrt{H}$ dependence, indicative of linear dispersion of the energy spectrum at the Fermi level in the superconducting state.
This behavior can be observed in the polar-angle dependence of $C_p$ shown in Fig.~\ref{fig2}(d).
The dip-type behavior toward the parallel direction (0$^{\circ}$) can be described by the following formula based on the anisotropic effective mass model:
\begin{equation}
\begin{split}
C_{p} (\theta)=C_{p}(0^{\circ})+\Delta C_{p}[1-(\sqrt{{\rm cos}^{2}\theta+\epsilon^{2}{\rm sin}^{2}\theta})^{n}]/(1-\epsilon^{n}),\\
\Delta C_{p}=C_{p}(90^{\circ})-C_{p}(0^{\circ}).\end{split}
\end{equation}
In this model, we assume that the angle dependence originates from the anisotropy of the critical field $H_{c2}$ as expressed by the anisotropic factor $\epsilon$.
In the case of organic superconductors, the paramagnetic pair-breaking effect is almost isotropic, and therefore, the origin of the anisotropy can be attributed to the orbital pair-breaking effect.
This means that the anisotropic factor $\epsilon$ can be described as $\epsilon$=$H_{\rm orb\parallel}$$/$$H_{\rm orb\perp}$=$\sqrt{m_{\perp}/m_{\parallel}}$, where $H_{\rm orb}$ and $m$ are the orbital limit and the effective mass, respectively.
The obtained values of $\epsilon$ about 10 ($m_{\perp}$/$m_{\parallel}$$\sim$100) are an indication of the two-dimensionality of these salts.
The power index $n$ reflects the low-energy excitation, which corresponds to the magnetic field dependence of the electronic heat capacity, and therefore, $n$=1/2 exactly agrees with the observed $\sqrt{H}$ dependence.
The $\sqrt{H}$ dependence and the value $n$=1/2 are consistent with the quasiparticle excitation observed as the $T^2$ dependence.
It should be noticed that $\kappa$-Br exhibits non-monotonic field dependence (a kink around $H$/$H_{\rm c2}$$\sim$0.5) different from that of $\kappa$-Ag as described by the thick curves that are guides for the eye.
The different behaviors between the two imply the difference in the gap structure.

\begin{figure}
\begin{center}
\includegraphics[width=\hsize,clip]{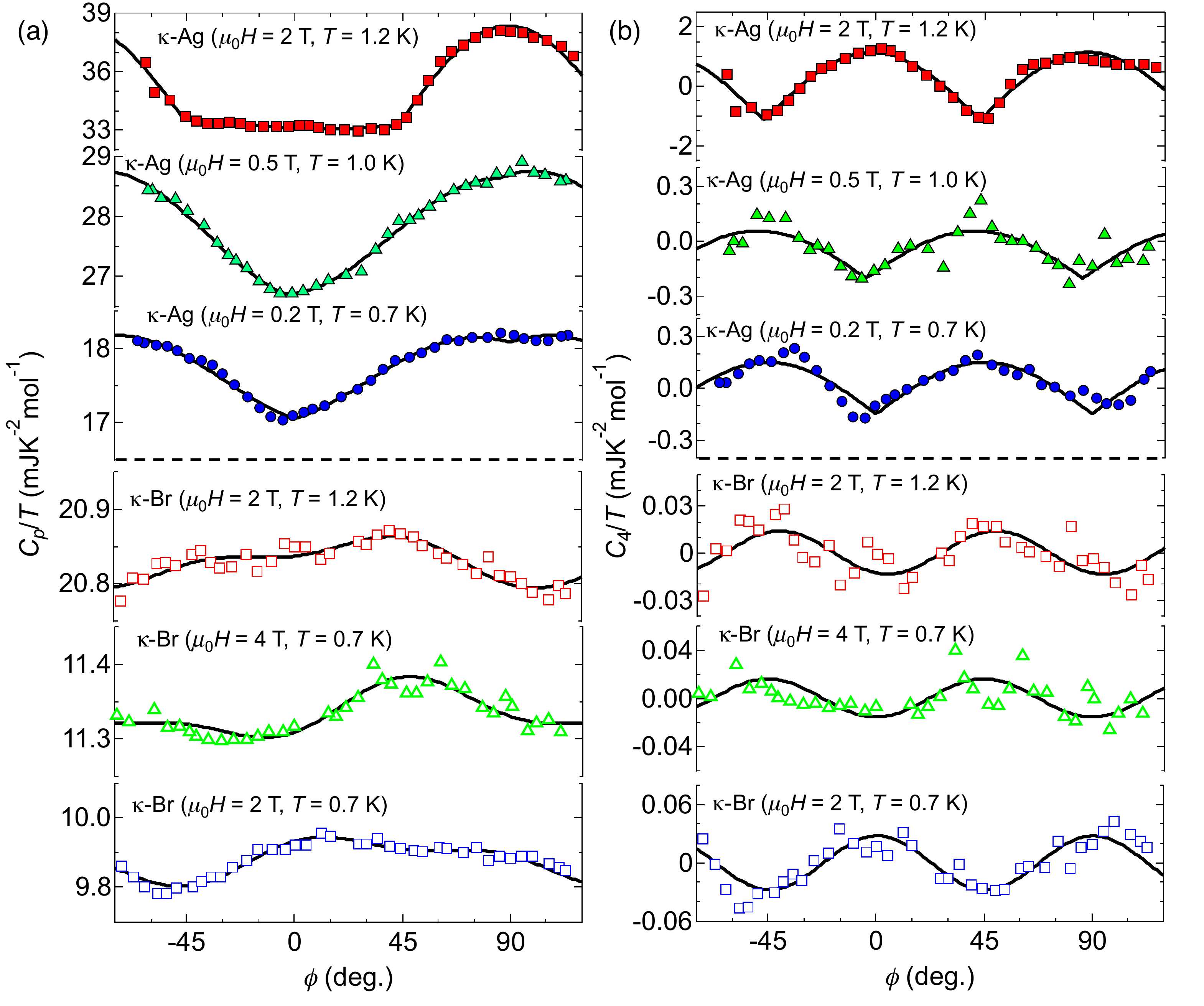}
\end{center}
\caption{
(a) Raw data of in-plane field-angle dependence of the heat capacity.
The black curves are fits to the formula introduced in the text.
(b) The fourfold oscillatory component derived from the data shown in (a) by subtracting the other components.
The black curves for $\kappa$-Ag (top panel) are fits to $C_4$=$C_4^{\prime}\left|{\rm cos}(2\phi)\right|$, while those for $\kappa$-Br (bottom panel) are fits to $C_4$=$C_4^{\prime}$cos(4$\phi$).
Since the $C_4$ term should not be expressed by such simple forms, the fittings are just approximations to describe the observed fourfold symmetry, leading to the difference of the fitting functions.
}
\label{fig3}
\end{figure}

\begin{figure}
\begin{center}
\includegraphics[width=\hsize,clip]{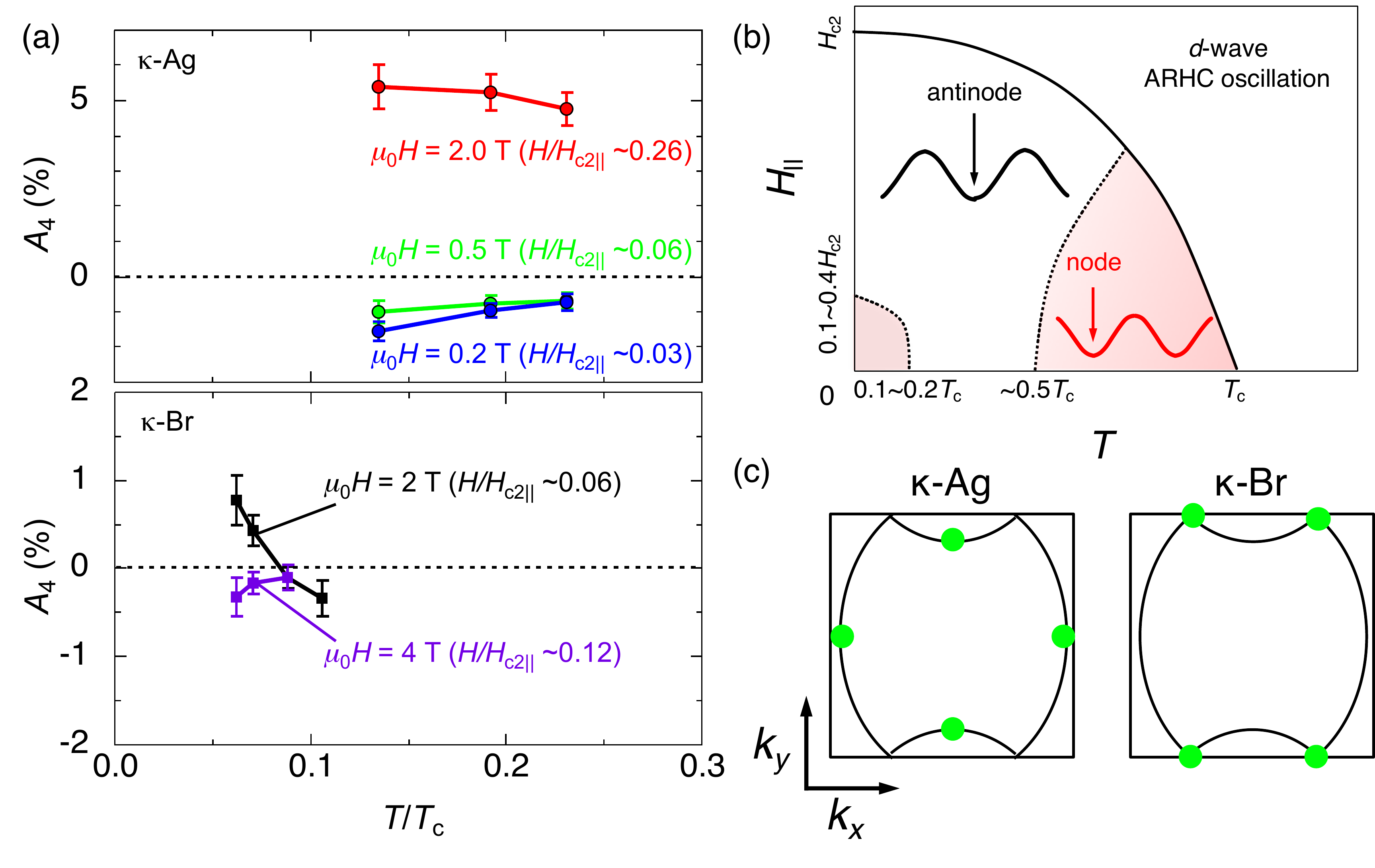}
\end{center}
\caption{
(a) The amplitude of the fourfold oscillatory term in various fields as a function of reduced temperature $T$/$T_{\rm c}$.
The positive (negative) value indicates that the maxima (minima) of the $C_4$ term are located at the directions of the crystal axes.
(b) Schematic phase diagram of angle-resolved heat capacity for $d$-wave superconductivity.
In the red areas, the node positions are located at the directions of the minima of angle-resolved heat capacity.
In the low-energy limit, namely, the low-$T$ and low-$H$ region, the node location corresponds to the minima of the $C_4$ term due to the predominant zero-energy Doppler effect\cite{19,20,21}.
The dotted curves indicate the region at which the sign of the $C_4$ term reverses and the amplitude of that becomes almost zero.
(c) Determined positions of the gap nodes in the Fermi surface for $\kappa$-Ag and $\kappa$-Br.
The green dots signify the position of the nodes.
}
\label{fig4}
\end{figure}

 To examine the nodal structures, we discuss the in-plane anisotropy for determining the positions of the line nodes.
In Fig.~\ref{fig3}(a), the heat capacity at various conditions as a function of the azimuthal angle $\phi$ from the $a$-axis direction is displayed.
The data indicate that the anisotropy has twofold and fourfold components, as reproduced by $C_p$($\phi$)/$T$=$C_0$/$T$+$C_2$/$T$+$C_4$/$T$, where $C_2$ and $C_4$ indicate the twofold and fourfold terms.
Although the anisotropy may not be expressed in a simple sinusoidal form, here we simply use the equations $C_2$=$C_{2}^{\prime}$cos[2($\phi$-$\phi$$_2$)] and $C_4$=$C_4^{\prime}\left|{\rm cos}(2\phi)\right|$ or $C_4$=$C_4^{\prime}$cos(4$\phi$) to evaluate the amplitude of the components.
The origin of $C_4$ can be assigned to the anisotropy of the $d$-wave gap function because the crystal structure does not have fourfold rotational symmetry.
In Fig.~\ref{fig3}(b), we extract the fourfold components $C_4$/$T$ from the total heat capacity by subtracting the other terms.
It is obviously seen that the amplitude and sign of the fourfold terms depend on temperature and magnetic field.
Figure~\ref{fig4}(a) displays the thermal variation of the percentage of the fourfold amplitude $A_4$=$C_4^{\prime}$/[$C_p$($\mu$$_0$$H$)$-$$C_p$(0~T)], scaled by the field-dependent parts of $C_p$ at each temperature.
The theories and experiments for angle-resolved heat capacity measurements\cite{19,20,21} suggest that the Doppler effect for the $d$-wave superconductivity gives a fourfold oscillation in the field-angle dependence of the density of states.
When considering only the zero-energy density of states, the minima of the fourfold term is parallel to the directions of the gap nodes.
Nevertheless, we need to take care that the oscillations show the sign change depending on temperature and field due to the contribution of the finite-energy density of states.
The correspondence of the positions between the gap nodes and the oscillation minima holds in the low-temperature and low-field region, as schematically shown by the red area in Fig.~\ref{fig4}(b)\cite{19,21}.
At the dotted curves, the sign of the $C_4$ term reverses due to the contribution of the finite-energy density of states.
Even though the size and shape of the area in which the zero-energy Doppler effect is predominant are greatly influenced by the structure of the Fermi surface\cite{20}, the positions of the gap nodes can be determined by the sign in the low-energy limit when the sign crossover is observed.
Thus, the negative sign at $\mu$$_0$$H$=0.2~T (=0.03$H_{c2\parallel}$), $T$=0.7~K (=0.13$T_{\rm c}$) for $\kappa$-Ag and the positive sign at $\mu$$_0$$H$=2~T (=0.06$H_{c2\parallel}$), $T$=0.7~K (=0.07$T_{\rm c}$) for $\kappa$-Br indicate that the gap nodes of $\kappa$-Ag are located at the $k_x$ and $k_y$ directions whereas those of $\kappa$-Br are positioned at $\phi$=45$^{\circ}$ apart from the crystal axes, as schematically depicted in Fig.~\ref{fig4}(c).
These positions mean that their fourfold symmetry is assigned to $d_{xy}$-wave for $\kappa$-Ag and $d_{x^{2}-y^{2}}$-wave for $\kappa$-Br.

 Next, we give consideration to the origin of the twofold term $C_2$.
Detailed analyses of this $C_2$ component are shown in Supplemental Material.\cite{Supp}.
If the two-dimensional plane of the measured sample was slightly misaligned from the magnetic field, the quasiparticle excitation by perpendicular fields also contributes to the angle dependence.
However, the form of this contribution in these salts should be $\sqrt{\left|{\rm cos}(\phi)\right|}$, different from the observed $C_{2}^{\prime}$cos[2($\phi$-$\phi$$_2$)].
This difference indicates that the twofold term is intrinsic to the present salts.
Since the point group of $\kappa$-Br is orthorhombic $D_{2h}$, the $d_{x^{2}-y^{2}}$ symmetry is categorized in the $A_{1g}$ irreducible representation, and the $s$-wave symmetry also belongs to the same representation in this point group.
This leads to the linear combination of the $s$-wave and $d_{x^{2}-y^{2}}$-wave symmetries in the $A_{1g}$ symmetry as the $d_{x^{2}-y^{2}}$+$s_{\pm}$-wave symmetry.
This situation is different from the $d_{xy}$ symmetry ($B_g$) for the monoclinic $C_{2h}$ group of $\kappa$-Ag.
Some recent studies\cite{8,16,23} point out that the $d_{x^{2}-y^{2}}$+$s_{\pm}$-wave gap function possesses twofold rotational symmetry due to the mixing of the $s_{\pm}$-wave gap.
The $C_2$ term observed by experiments may be the symmetry origin in the case of $\kappa$-Br.
However, $\kappa$-Ag, namely $d_{xy}$-wave superconductivity, also shows a clear $C_2$ component.
In other angle-resolved measurements for the dime-Mott superconductors\cite{5,9}, the origin has been discussed in terms of the anisotropy of the ellipsoidal Fermi surface because the anisotropy of the Fermi velocity also contributes to the quasiparticle excitation outside the vortices\cite{20,21}.
We therefore consider that the appearance of the $C_2$ term observed in both compounds originates from the combination of the anisotropy of the gap function and the Fermi velocity and it should be a common feature for organic superconductors with ellipsoidal Fermi surfaces.

 Our results elucidate the symmetries of $\kappa$-Ag and $\kappa$-Br as $d_{xy}$ and $d_{x^{2}-y^{2}}$+$s_{\pm}$, respectively, even though they are classified into the same $\kappa$-type dimer-Mott system.
Why do these compounds have different symmetries?
To answer the question, we here compare the present results with the theoretical studies for the pairing mechanisms based on the antiferromagnetic spin fluctuations.
In fact, recent theoretical calculations without the dimer approximation\cite{15,16,17,18} pointed out that the $d_{xy}$- and $d_{x^{2}-y^{2}}$+$s_{\pm}$-wave symmetries compete according to the transfer integrals in the dimer units and the geometric frustration of the dimer triangular lattice.
Even though Guterding, $et al$.\cite{16} suggested that $d_{x^{2}-y^{2}}$+$s_{\pm}$-wave symmetry is stable for all of the discovered $\kappa$-type salts, it should be noticed that $\kappa$-Ag is located at the verge of the phase boundary between $d_{xy}$ and $d_{x^{2}-y^{2}}$+$s_{\pm}$ in the diagram in their calculation.
Therefore, it is possibly reasonable that the superconducting symmetry of $\kappa$-Ag is $d_{xy}$ because of some differences between the theories and our experiments, such as the temperature dependence of the transfer integrals.
Moreover, we should notice that $\kappa$-Br is also near the boundary even though the parameters are different.
The previous work on the angle-resolved heat capacity by Malone, $et al$.\cite{9} indicated that the symmetry of $\kappa$-Br is attributed to $d_{xy}$.
This apparent disagreement could also come from the proximity to the boundary of the two symmetries since $T_c$ of our salt ($\sim$11.3~K), lower than that of Ref.~\cite{9} ($\sim$12.5~K), indicates the difference of the condition, such as a small applied pressure in the measurement, which slightly modifies the effective transfer integrals.
Considering these factors, we can claim that the superconductivity is mediated by antiferromagnetic spin fluctuations, but the two types of symmetry are nearly degenerate and the lattice geometry finally determines the emergent symmetry according to the balance of the transfer integrals.

 In summary, we examine the symmetry of the superconductivity in $\kappa$-(BEDT-TTF)$_2$$X$ ($X$=Cu[N(CN)$_2$]Br and $X$=Ag(CN)$_2$H$_2$O) by means of the heat capacity measurements.
The $T^2$ temperature and $\sqrt{H}$ field dependence of the heat capacity in the low-energy region indicates the presence of line nodes in the Fermi surface.
The scenario is certainly consistent with the reported results based on two-dimensional $d$-wave superconductivity.
From the results of the angle-resolved heat capacity measurements, we established the possible pairing symmetry of the superconductivity as $d_{xy}$ for $\kappa$-Ag and $d_{x^{2}-y^{2}}$+$s_{\pm}$ for $\kappa$-Br.
Comparing our results with the theoretical studies, we find that the apparently contradictory result can be rationalized by the ratio of the transfer integrals between the dimers.
This indicates that the symmetry of $\kappa$-(BEDT-TTF)$_2$$X$ is dominated by the strength of the dimerization and the distorted triangular dimer lattice even though the superconductivity is caused by the same pairing mechanism, the antiferromagnetic spin fluctuations.

 We thank K. Kawamura (Osaka University) and Dr. H. Akutsu (Osaka University) for kind support in the x-ray structural analysis.

\renewcommand{\thefigure}{S\arabic{figure}}
\clearpage
\begin{center}
\large{\bf{Supplemental Material for\\
Symmetry change of $d$-wave superconductivity\\
in $\kappa$-type organic superconductors
}}
\end{center}
\section{Twofold term $C_2$ in the in-plane anisotropy}
\begin{figure}
\begin{center}
\includegraphics[width=\hsize,clip]{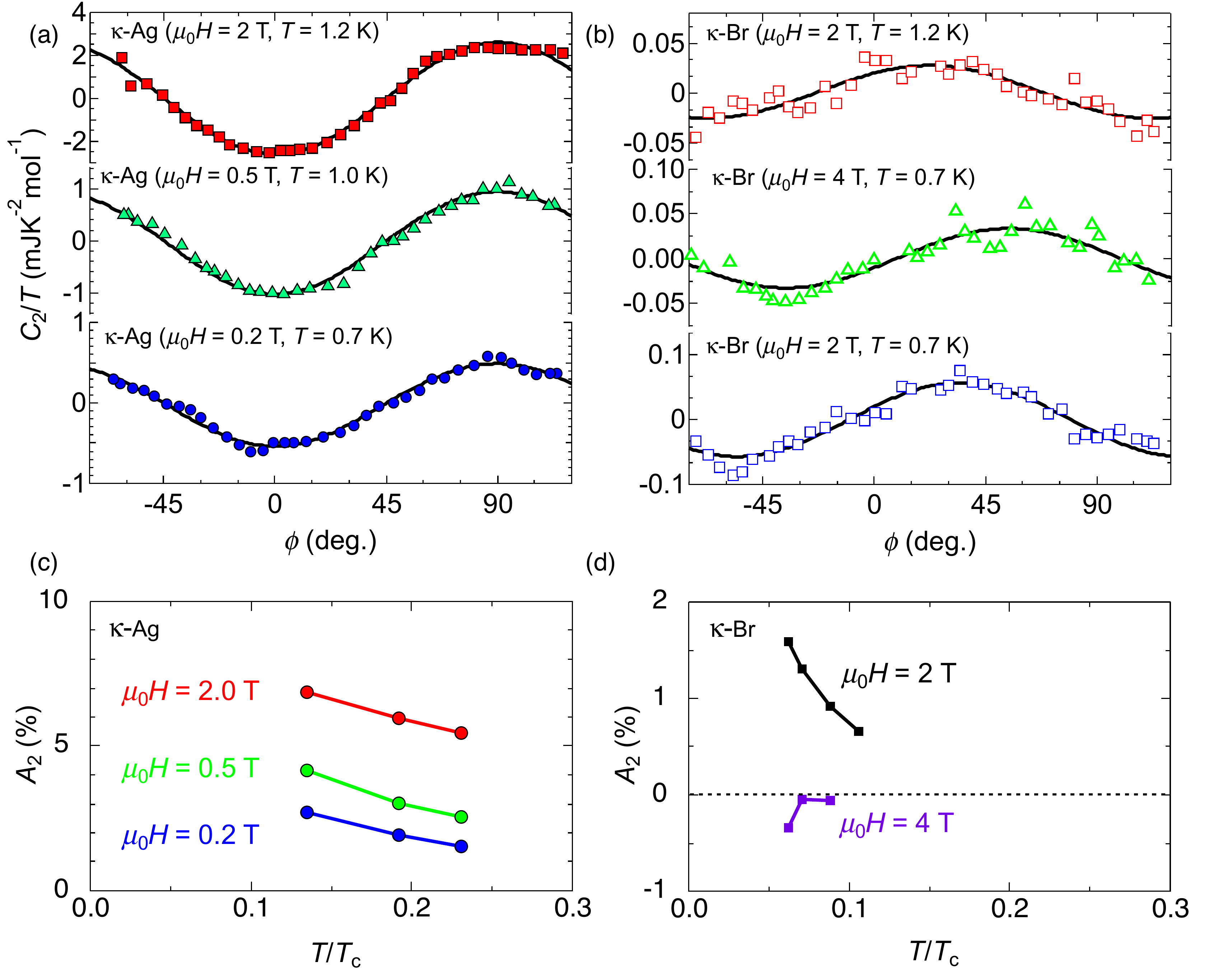}
\end{center}
\caption{
(a),(b) Twofold oscillatory component derived from the data shown in Fig.~3(a) (the main text) by subtracting the other components.
The solid curves are the fits to the formula $C_2$/$T$=$C_{2}^{\prime}$cos[2($\phi$-$\phi$$_2$)].
(c),(d) $A_2$ as a function of reduced temperature $T$/$T_{\rm c}$.
}
\label{figS1}
\end{figure}

In this section, we present the twofold term $C_2$, extracted from the data shown in Fig.~3(a).
Fig.~\ref{figS1} displays the temperature and field dependence of the twofold components $C_2$/$T$, extracted from the total heat capacity by subtracting the other terms.
As discussed in the main text, if $C_2$ originates from the sample misalignment against magnetic fields, it can be described by the formula $\sqrt{\left|{\rm cos}(\phi)\right|}$.
However, the observed twofold term $C_2$ can be a simple twofold sinusoidal fit $C_{2}^{\prime}$cos[2($\phi$-$\phi$$_2$)].
This means that the present twofold component arises from the in-plane anisotropy of the salts.
To evaluate the twofold contribution, we here scale this component with the change of the electronic heat capacity as $A_2$=$C_2^{\prime}$/[$C_p$($\mu$$_0$$H$)$-$$C_p$(0~T)].
In Fig.~\ref{figS1}(c), we find that the temperature-dependent behavior can be roughly scaled by $\sqrt{H}$ for $\kappa$-Ag without any phase shift, which indicates that the $C_2$ term simply reflects the quasiparticles excited by magnetic fields.
In contrast, $\kappa$-Br shows the phase shift (Fig.~\ref{figS1}(b)) depending on field, which leads to the change of the sing as shown in Fig.~\ref{figS1}(d).
The phase shift of the twofold term indicates that the $C_2$ term should be composed of some elements, such as the anisotropy of the Fermi velocity and the twofold rotational symmetry of the $d_{x^{2}-y^{2}}$+$s_{\pm}$-wave gap function.
Indeed, the behavior of $\kappa$-Br is roughly similar to the temperature and field dependence of the $C_4$ term coming from the gap symmetry, which is different from the case of $\kappa$-Ag.
The difficulty to decompose these components makes the detailed discussion hard.
Nevertheless, this fact implies that the $C_2$ term of $\kappa$-Ag comes only from the anisotropy of the Fermi velocity, whereas that of $\kappa$-Br may result from not only the anisotropy of the Fermi velocity but also that of the $d_{x^{2}-y^{2}}$+$s_{\pm}$-wave gap symmetry.
This possible viewpoint agrees with the discussion on the $C_4$ term.


\begin{thebibliography}{99}
\bibitem{4.5} From the viewpoint of the point group, $d_{x^{2}-y^{2}}$-wave symmetry is categorized into $A_{1g}$ for the $D_{2h}$ point group and $A_g$ for $C_{2h}$, which also contain $s$-wave symmetry. Thus, in the case of $\kappa$-(BEDT-TTF)$_2$X salts, the $d_{x^{2}-y^{2}}$ wave is equivalent to $d_{x^{2}-y^{2}}$+$s_{\pm}$ in terms of symmetry classification.
\bibitem{29.5} The Pauli-limiting field $H_{\rm P\parallel}$ is assumed to be the in-plane $H_{\rm c2\parallel}$ here because it is regarded as the critical field of the uniform superconductivity if the possible FFLO state does not occur above $H_{\rm P\parallel}$.
\end{thebibliography}
\end{document}